\def\cA{{\cal A}}

\def\plushc{\ + h.c.}

\newcommand{\I}{{\rm i}}
\newcommand{\e}{{\rm e}}

\newcommand{\thet}{\theta}
\newcommand{\diag}{{\rm diag}}
\newcommand{\cL}{{\cal L}}
\newcommand{\cD}{{\cal D}}
\newcommand{\cF}{{\cal F}}

\newcommand{\nn}{\nonumber}
\newcommand{\al}[1]{\begin{eqnarray}#1\end{eqnarray}}
\newcommand{\eq}[1]{\begin{equation}#1\end{equation}}

\def\be{\begin{equation}}
\def\ee{\end{equation}}
\def\bea{\begin{eqnarray}}
\def\eea{\end{eqnarray}}
\def\Ref#1{(\ref{#1})}
\def\cropen#1{\crcr\noalign{\vskip #1}}
\def\crr{\cropen{1\jot }}
\def\Red{}
\def\Black{}
\documentstyle[epsf, 12pt]{article}
\begin{document}

\thispagestyle{empty}
\begin{flushright}
JHU-TIPAC-99008\\
hep-th/9908084\\
\end{flushright}

\bigskip\bigskip\begin{center} {\bf
\LARGE{\bf Dual Anti-de Sitter Superalgebras from Partial Supersymmetry
Breaking} }
\end{center} \vskip 1.0truecm

\centerline{\bf Richard Altendorfer and Jonathan Bagger}

\vskip5mm
\centerline{\it Department of Physics and Astronomy}
\centerline{\it The Johns Hopkins University}
\centerline{\it 3400 North Charles Street}
\centerline{\it Baltimore, MD 21218, U.S.A.}
\vskip5mm

\bigskip \nopagebreak \begin{abstract}
\noindent The partial breaking of supersymmetry in anti-de Sitter
(AdS) space can be accomplished using two of four dual representations
for the massive $OSp(1,4)$ spin-3/2 multiplet. The representations
can be ``unHiggsed,'' which gives rise to a set of dual $N=2$
supergravities and supersymmetry algebras.
\end{abstract}

\newpage\setcounter{page}1

\section{Introduction}

Most supersymmetry phenomenology is based on the minimal
supersymmetric standard model, which is assumed to be
the low energy limit of a more fundamental theory in which
an $N=8$ supersymmetry is spontaneously broken to $N=1$.
The Goldstone fermions from the seven broken supersymmetries
are eliminated by the superHiggs effect: they become the
longitudinal components of seven massive gravitinos.

The physics that underlies this superHiggs effect is most
easily illustrated by the simpler case of  $N=2$ supersymmetry,
spontaneously broken to $N=1$.  Indeed, in Ref.~\cite{np} a
set of theories were constructed that realized this partial
supersymmetry breaking in flat Minkowski space.  These theories
circumvent a naive argument that forbids partial breaking by
abandoning one of its assumptions, namely that of a positive
definite Hilbert space.  In covariantly-quantized supergravity
theories, the gravitino $\psi_{m\alpha}$ is a gauge field with
negative-norm components, so the Hilbert space is not positive
definite.

The Minkowski-space theories were based on $N=2$
super-Poincar\'e algebras with certain central extensions.
In anti-de Sitter (AdS) space, however, the $N=2$
supersymmetry algebra is different.  The algebra
is known as $OSp(2,4)$; the relevant parts are (see
appendix {\bf A}),
\bea
\{ Q^i_\alpha, \bar Q_{j\dot\beta}\}&\ =\ &
2\sigma^a_{\alpha\dot\beta}R_a\delta^i_j
\nonumber \\
\{ Q^i_\alpha, Q^{\beta j}\}&=&2\I \Lambda
{{\sigma^{ab}}_\alpha}^\beta M_{ab}\delta^{ij} +
 2\I {\delta_\alpha}^\beta T^{ij}
\nonumber \\[1mm]
\left[ T^{ij}, Q^k \right]&=&\I \Lambda
(\delta^{jk}Q^i - \delta^{ik}Q^j)\ .
\eea
In this expression, the $Q^i_\alpha$ $(i \in \{1,2\})$
denote the two supercharges, while $M_{ab}$ and $R_a$
are the generators of $SO(3,2)$. The antisymmetric
matrix $T^{ij}$ is the single hermitian generator of
an additional $SO(2)$.  As the cosmological constant
$\Lambda \rightarrow 0$, the algebra contracts to the
usual $N=2$ Poincar\'e supersymmetry algebra with at
most one real central charge.  (The generator $R_a$
contracts to the momentum generator $P_a$, while
$T^{ij}$ contracts to zero or to a single real central
charge, depending on the rescaling of the operators.)

In Minkowski space, partial supersymmetry breaking
was found to require super-Poincar\'e algebras with
{\it two} central extensions \cite{np}.  The $OSp(2,4)$
algebra contracts to a super-Poincar\'e algebra with
at most {\it one} central charge.  This suggests that
if partial breaking is to occur, the AdS algebra must
be modified.

In this paper we will study this question using
the same approach as in \cite{np}.  We will see
that partial breaking in AdS space occurs for
two of four dual representations of the $OSp(1,4)$
massive spin-3/2 multiplet.  We will find that
the two dual representations give rise to new
AdS supergravities with appropriately modified
$OSp(2,4)$ supersymmetry algebras\footnote{A theory
exhibiting partial supersymmetry breaking in
AdS space was derived in \cite{cgpads}.  However,
this construction contains more fields because
it has complete $N=2$ multiplets.}.  As the
cosmological constant $\Lambda \rightarrow 0$,
the new algebras contract to the $N=2$ Poincar\'e
algebras with the required set of central
extensions.

\section{The SuperHiggs Effect in Partially Broken
AdS Supersymmetry}

\subsection{Dual Versions of Massive $OSp(1,4)$
Spin-3/2 Multiplets}

The starting point for our investigation is the
massive $OSp(1,4)$ spin-3/2 multiplet.  This
multiplet contains six bosonic and six fermionic
degrees of freedom, arranged in states of the
following spins,
\be
\pmatrix{
{3\over2} \crr
1\ \ \ 1 \crr
{1\over2}}\ .
\ee
It contains the following AdS representations (see
e.g.~\cite{nicolai} and references therein),
\be
D(E+{\textstyle{1\over2}},
{\textstyle{3\over2}})\ \oplus\ D(E,1)
\ \oplus \
D(E+1,1)\ \oplus\
D(E+{\textstyle{1\over2}},
{\textstyle{1\over2}})
\label{reps}
\ee
where $D(E,s)$ is labeled by the eigenvalues of the diagonal operators
of the maximal compact subgroup 
$ SO(2) \times SU(2) \subset
SO(3,2)$ and
unitarity requires $E \geq 2$.  (The eigenvalue
$E$ is the AdS generalization of a representation's
rest-frame energy.   As $E \rightarrow 2$, the
first two representations in \Ref{reps} become
``massless," with eigenvalues $(s+1, s)$.  The
massless representations are short representations
of $OSp(1,4)$.)

As in Minkowski space, a massive spin-1 field can
be represented by a vector or by an antisymmetric
tensor.  For the case at hand, there are four
possibilities.  The Lagrangian with two vectors
is given by
\bea
e^{-1}\cL &\ =\ & e^{-1}\epsilon^{m n r s} \overline \psi_{m}
  \overline \sigma_n \nabla_r \psi_s
 - \I \overline \zeta \overline \sigma^m \nabla_m \zeta
 - {1 \over 4} A_{m n} A^{m n} - {1 \over 4} B_{m n} B^{m n}
\nonumber \\ & & -\ {1\over 2}(m^2 - m\Lambda) A_m A^m
 - {1\over 2}(m^2 + m\Lambda) B_m B^m  \nonumber \\
& & +\ {1\over 2}m\,\zeta\zeta \ +\ {1\over 2}m\,\bar\zeta\bar\zeta
 -\ m\,\psi_m \sigma^{m n} \psi_n \ -\ m\,\bar\psi_m \bar\sigma^{m n}
\bar\psi_n
\label{AdSL}
\eea
where $\Lambda \geq 0$ and $\nabla_m$ is the AdS
covariant derivative.  Here $\psi_m$ is a spin-3/2
Rarita-Schwinger field, $\zeta$ a spin-1/2 fermion,
and $A_{mn}$ and $B_{mn}$ are the field strengths
of the real vectors $A_m$ and $B_m$.  This Lagrangian
is invariant under the following supersymmetry
transformations,\footnote{Here, and in all subsequent
rigid supersymmetry transformations, the parameter
$\eta$ is covariantly constant but $x$-dependent
(see \Ref{xdep} in appendix {\bf A}).}
\bea
\delta_\eta A_m &=& \sqrt{1 + \epsilon}(\psi_m\eta +
\bar\psi_m\bar\eta) \nonumber \\
& & +\ {1\over\sqrt{1 - \epsilon}}
\left( \I{1\over\sqrt{3}}(1 - \epsilon)
(\bar\eta\bar\sigma_m\zeta - \bar\zeta\bar\sigma_m\eta)
-{1\over \sqrt{3}m}\partial_m(\zeta\eta + \bar\zeta\bar\eta)\right) \nonumber
\\
\delta_\eta B_m &=& \sqrt{1 - \epsilon}(-\I\psi_m\eta + \I\bar\psi_m\bar\eta)
\nonumber \\
& & +\ {1\over\sqrt{1 + \epsilon}}\left( - {1\over\sqrt{3}}(1 +
\epsilon) (\bar\eta\bar\sigma_m\zeta + \bar\zeta\bar\sigma_m\eta)
+{\I\over \sqrt{3}m}\partial_m(\zeta\eta - \bar\zeta\bar\eta) \right)
\nonumber\\
\delta_\eta \zeta &=& \sqrt{1 -
\epsilon}({1\over\sqrt{3}}A_{mn}\sigma^{mn}\eta -\I{m\over\sqrt{3}}
\sigma^m\bar\eta A_m) \nonumber \\
& & +\ \sqrt{1 + \epsilon}( - {\I\over\sqrt{3}}B_{mn}\sigma^{mn}\eta
+ {m\over\sqrt{3}}
\sigma^m\bar\eta B_m ) \nonumber \\
\delta_\eta \psi_m &=& {1\over\sqrt{1 + \epsilon}} \left( {1\over
3m}\nabla_m(A_{rs}\sigma^{rs}\eta + 2\I m
\sigma^n\bar\eta A_n)
 -{\I\over 2}(H^A_{+mn}\sigma^n + {1\over
3}H^A_{-mn}\sigma^n)\bar\eta \right. \nonumber \\ & & \left.
-\ {2\over 3}m({\sigma_m}^n A_n \eta + A_m\eta)
 -{\I\over 2}\epsilon H^A_{+mn}\sigma^n\bar\eta - \epsilon m
A_m\eta \right) \nonumber \\ & & +\ {1\over\sqrt{1 - \epsilon}}\left(
{-\I\over 3m}\nabla_m(B_{rs}\sigma^{rs}\eta - 2\I m
\sigma^n\bar\eta B_n)
    +{1\over 2}(H^B_{+mn}\sigma^n \right. \nn\\
    &&  +\ {1\over
3}H^B_{-mn}\sigma^n)\bar\eta  \left.
+ {2\over 3}\I m({\sigma_m}^n B_n \eta + B_m\eta)
    -{1\over 2}\epsilon H^B_{+mn}\sigma^n\bar\eta - \I\epsilon m
B_m\eta \right) \ ,\nn\\
\eea
where $H^A_{\pm mn}=A_{mn}\pm {\I\over 2}\epsilon_{mnrs}A^{rs}$
and $\epsilon = \Lambda/m$.  Note that the ``mass'' $m$ is defined
to be $m = (E-1)\Lambda$.  This definition is consistent with
the AdS representations in (\ref{reps}). The fact that $E
\geq 2$ implies that $0 \le \epsilon\le 1$.

In Minkowki space, other field representations of the massive spin-3/2 multiplet 
can be derived
\cite{np} using a Poincar\'e duality which relates massive
vector fields to massive antisymmetric tensor fields of rank
two.  The same duality also holds in AdS space where, for
example, the vector $B_m$ can be replaced by an antisymmetric
tensor $B_{mn}$.  The Lagrangian for the dual theory is
then:
\al{
e^{-1}\cL &\ =\ & e^{-1}\epsilon^{m n r s} \overline \psi_{m}
  \overline \sigma_n \nabla_r \psi_s
 - \I \overline \zeta \overline \sigma^m \nabla_m \zeta
 - {1 \over 4} A_{m n} A^{m n} + {1\over 2}v^{Bm} v^B_m
\nonumber \\ & &
 - {1\over 2}(m^2 - m\Lambda) A_m A^m -\ {1\over 4}(m^2 + m\Lambda)
B_{mn}B^{mn} \nonumber \\
& & +\ {1\over 2}m\,\zeta\zeta \ +\ {1\over 2}m\,\bar\zeta\bar\zeta
 -\ m\,\psi_m \sigma^{m n} \psi_n \ -\ m\,\bar\psi_m \bar\sigma^{m n}
\bar\psi_n
}
where $A_{mn}$ is the field strength associated with the real
vector field $A_m$ and $v_m$ is the field strength for the
antisymmetric tensor $B_{mn}$.  This Lagrangian is invariant
under the following supersymmetry
transformations:\footnote{Here, the square brackets denote
antisymmetrization, without a factor of 1/2.}
\al{
\delta_\eta A_m &=& \sqrt{1 + \epsilon}(\psi_m\eta + \bar\psi_m\bar\eta)
\nonumber \\
& & +\ {1\over\sqrt{1 - \epsilon}}
\left( \I{1\over\sqrt{3}}(1 - \epsilon)
(\bar\eta\bar\sigma_m\zeta - \bar\zeta\bar\sigma_m\eta)
-{1\over \sqrt{3}m}\partial_m(\zeta\eta + \bar\zeta\bar\eta)\right) \nonumber
\\
\delta_\eta B_{mn} &=& \sqrt{1 - \epsilon\over 1 + \epsilon}
 \left(- {1\over m}\nabla_{[ m}(\eta\psi_{n ]}) - \I \eta\sigma_{[ m}\bar\psi_{n ]}
\right)
- {2\over\sqrt{3}}\left( \bar\eta\bar\sigma_{mn}\bar\zeta
+ {\I\over 2m}\nabla_{[m}(\bar\zeta\bar\sigma_{n]}\eta) \right) \plushc \nn \\
\delta_\eta \zeta &=& \sqrt{1 -
\epsilon}({1\over\sqrt{3}}A_{mn}\sigma^{mn}\eta -\I{m\over\sqrt{3}}
\sigma^m\bar\eta A_m) \nonumber \\
& & +\  {m\over\sqrt{3}}(1 + \epsilon)B_{mn}\sigma^{mn}\eta
+ {1\over\sqrt{3}}
\sigma^m\bar\eta v^B_m  \nonumber \\
\delta_\eta \psi_m &=&
{1\over\sqrt{1 + \epsilon}} \left( {1\over
3m}\nabla_m(A_{rs}\sigma^{rs}\eta + 2\I m
\sigma^n\bar\eta A_n)
 -{\I\over 2}(H^A_{+mn}\sigma^n + {1\over
3}H^A_{-mn}\sigma^n)\bar\eta \right. \nonumber \\ & & \left.
-\ {2\over 3}m({\sigma_m}^n A_n \eta + A_m\eta)
 -{\I\over 2}\epsilon H^A_{+mn}\sigma^n\bar\eta - \epsilon m
A_m\eta \right) \nonumber \\
& & + {1\over\sqrt{1 - \epsilon}} \left( {1\over
3m}\nabla_m( m\sqrt{1 + \epsilon} B_{rs}\sigma^{rs}\eta - 2 {1\over\sqrt{1 +
\epsilon}}
\sigma^n\bar\eta v^B_n) \right. \nonumber \\
& & + \I m \sqrt{1 + \epsilon} (({1\over 3} - {\epsilon\over
2})B_{mn}\sigma^n\bar\eta + \I
({1\over 3} - {\epsilon\over 4})\epsilon_{mnrs}B^{n r}\sigma^s\bar\eta)
  \nonumber \\
& & \left. +\ {2\over 3}{\I\over\sqrt{1 + \epsilon}} ({\sigma_m}^n v^B_n \eta +
v^B_m\eta)
 - \I {\epsilon\over\sqrt{1 + \epsilon}}
v^B_m\eta \right) \ . \nonumber
}

Two more representations can be found by dualizing the
vector $A_m$.  The derivations are straightforward, so we
will not write the Lagrangians and transformations here.
Each of the four dual Lagrangians describe the dynamics
of free massive spin-3/2 and 1/2 fermions, together with
their supersymmetric partners, massive spin-one vector
and tensor fields.

In what follows we shall see that the first two
representations are special because they can be regarded
as ``unitary gauge'' descriptions of theories with a
set of additional symmetries:  a fermionic gauge symmetry
for the massive spin-3/2 fermion, as well as additional
gauge symmetries associated with the massive gauge
fields.

\subsection{UnHiggsing Massive $OSp(1,4)$ Spin-3/2
Multiplets}

To exhibit the superHiggs effect, we will unHiggs
these Lagrangians by first coupling them to $N=1$ supergravity and 
including a Goldstone fermion
and its superpartners, and then gauging the full $N=2$
supersymmetry.  In this way we will construct theories
with a local $N=2$ supersymmetry nonlinearly realized,
but with $N=1$ represented linearly on the fields.  The
resulting Lagrangians describe the physics of partial
supersymmetry breaking well below the scale $v$ where
the second supersymmetry is broken.

In flat space, a massive representation can be
unHiggsed by passing to its massless limit, where
the Goldstone fields become physical degrees of
freedom.  In AdS space, the ``massless" limit
corresponds to $E \rightarrow 2$.  In this limit
the massive spin-3/2 multiplet splits into a
massless spin-3/2 multiplet, plus a {\it massive}
vector/tensor multiplet of spin one (see also appendix {\bf B}):

\begin{figure}[ht]
\epsffile[60 575 490 720]{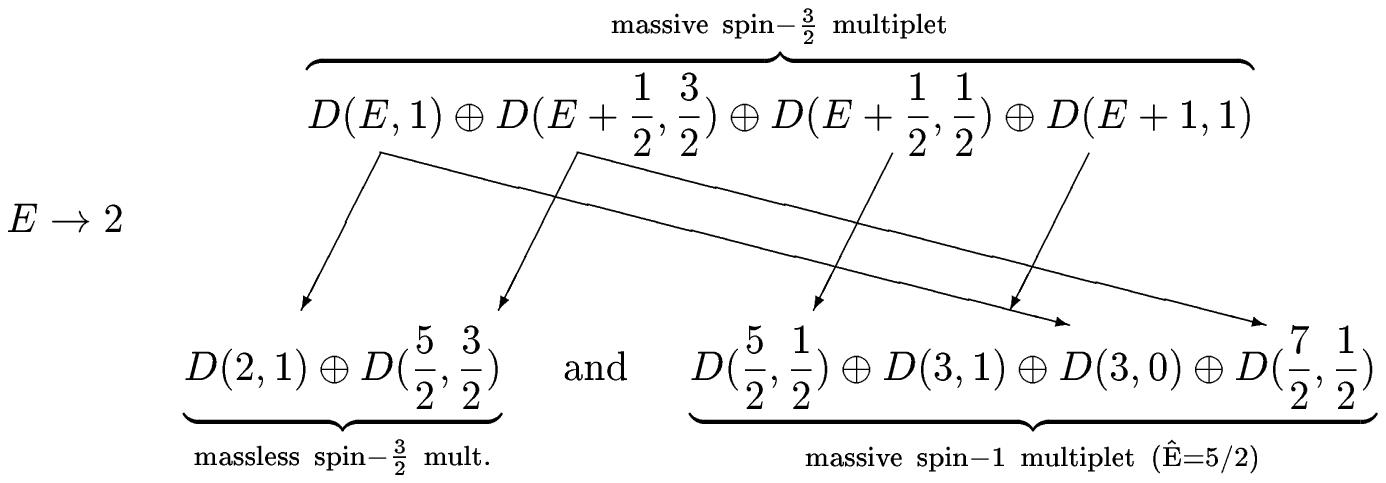}
\end{figure}

The spin-one multiplet with $\hat E=5/2$ (The normalization of $E$ differs for different multiplets;
see appendix {\bf B} and Ref.~\cite{nicolai}.) cannot itself be unHiggsed
because that would require $E \rightarrow 1$.  For
the case at hand, this would spoil the unitarity of
the spin-3/2 field.

In Figure 1 the physical fields of the 
massive spin-3/2 multiplet coupled to gravity are arranged in terms of 
$N=1$ multiplets.  The fields of lowest spin form a massive $N=1$ 
vector/tensor multiplet.  They may be thought 
of as $N=1$ ``matter.''  The remaining fields are the gauge fields of
$N=2$ supergravity.  In unitary gauge, the massless vector eats the
scalar, while the Rarita-Schwinger field eats one linear combination of 
the spin-1/2 fermions.  This leaves the massive $N=1$ spin-3/2 multiplet coupled 
to $N=1$ supergravity.
In contrast to the superHiggs effect in a flat background, where
an $N=2$ multiplet emerges in the massless
limit of the unHiggsed theory \cite{np}, the $E \rightarrow 2$ limit in AdS space
gives rise to $N=1$ multiplets only.

\begin{figure}[t]
\epsffile[70 575 375 745]{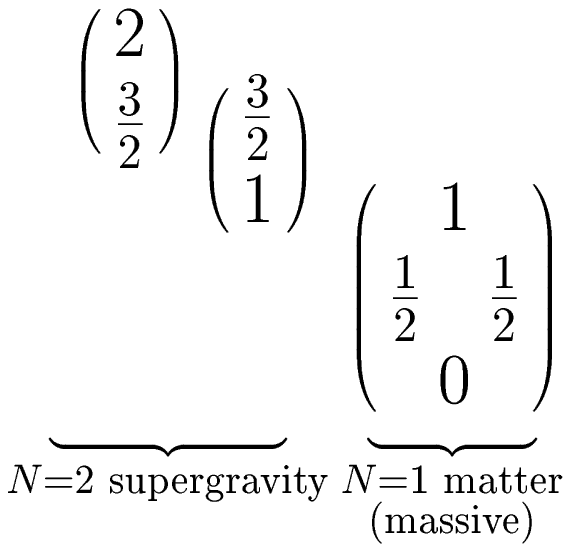}
\caption{The degrees of freedom of the unHiggsed $OSp(1,4)$ massive
spin-3/2 multiplet coupled to gravity. The massive spin-1 field can
be represented by either a vector or an antisymmetric tensor.}
\end{figure}

To find the Lagrangian, let us introduce a set of
Goldstone fields by the following St\"uckelberg
redefinitions.  For the case with two vectors,
we include Goldstone fields by replacing
\al{
A_{m} &\rightarrow& A_{m} - {1\over \sqrt{1 - \epsilon}m}\partial_m\phi_A
\nonumber \\
B_{m} &\rightarrow& B_{m} - {1\over \sqrt{1 + \epsilon}m}\partial_m\phi_B
\ . \label{st1}
}
For the dual representation, we take
\al{
A_{m} &\rightarrow& A_{m} - {1\over \sqrt{1 - \epsilon}m}\partial_m\phi
\nonumber \\
B_{mn}&\rightarrow& B_{mn} - {1\over \sqrt{1 + \epsilon}m}\partial_{[ m} B_{n
]}\ . \label{st2}
}
In each case, the introduction of the Goldstino $\nu$
requires an additional shift
\eq{
\psi_m\ \rightarrow\ \psi_m - {1\over\sqrt{6}\sqrt{1 - \epsilon^2}m}
(2 \nabla_m \nu + \I m \sigma_m \bar\nu) \label{st3}
}
to obtain a proper kinetic term for $\nu$.

For the case with two vectors, the Lagrangian is as follows,
\bea
& &e^{-1}\cL \ =\ \nonumber\\
&  & -\ {1 \over 2
\kappa^2} {\cal R} + \epsilon^{m n r
s}
\overline
\psi_{i m}
  \overline \sigma_n D_r \psi^i_s
 - \I \overline \lambda \overline \sigma^m D_m \lambda
 - \I \overline \chi \overline \sigma^m D_m \chi  \nonumber \\
& & -\ {1 \over 4} A_{m n} A^{m n} - {1 \over 4}
B_{m n} B^{m n} - {1\over 2}\cD_m \phi_A \cD^m
\phi_A
 - {1\over 2}\cD_m \phi_B \cD^m \phi_B   \nonumber \\
& & - \ \Bigl( {1\over \sqrt{2}} m \sqrt{1 - \epsilon^2}
\psi^2_m \sigma^m
   \overline \lambda
+ m \sqrt{1 - \epsilon^2}  \I\psi^2_m \sigma^m
   \overline \chi  \nonumber \\
& &+\ \sqrt{2} m \I\lambda\chi
   + {1\over 2}m\chi\chi
\Red +\ m\psi^2_m \sigma^{m n} \psi^2_n
-
     \epsilon m\psi^1_m \sigma^{m n} \psi^1_n \Black \nonumber \\
& &+\ {\kappa \over 4}  \epsilon_{i j} \psi^i_m
\psi_{n}^{j}
    (\sqrt{1 + \epsilon}H_{A-}^{m n} - \I \sqrt{1 - \epsilon}H_{B-}^{m
n})  \nonumber \\
& & +\ {\kappa \over {2}}  \chi \sigma^m
    \overline \sigma^n \psi^1_m (\cD_n \phi_A - \I \cD_n \phi_B)
 \nonumber \\
& &+\ {\kappa \over 2 \sqrt{2}}  \overline \lambda
    \overline \sigma_m \psi^1_n (\sqrt{1 - \epsilon}H_{A+}^{m n} - \I
\sqrt{1 + \epsilon}H_{B+}^{m n})
    \nonumber \\
& & +\  {\kappa \over {2}}
    \epsilon^{m n r s}\sqrt{1 - \epsilon\over 1 +
\epsilon}
    \overline \psi_{m 2} \overline \sigma_n \psi^1_r
    (\partial_s \phi_A - \I \partial_s
\phi_B)
    \nonumber \\
& & -\ {\kappa \over {2}}m
    \epsilon^{m n r s}
    \overline \psi_{m 2} \overline \sigma_n \psi^1_r
    (\sqrt{1 + \epsilon}A_s - \I\sqrt{1 -
\epsilon}B_s)
     \nonumber \\
& & -\ 2\kappa\epsilon m \sqrt{1 - \epsilon\over
1 + \epsilon} \bar\psi_{m 2}
\bar\sigma^{mn}\bar\psi_{n 1}\phi_A
+ {\kappa\epsilon m \over
\sqrt{2}}\bar\lambda\bar\sigma^m\psi^{1}_m \phi_A
  \nonumber \\
& & +\   \I {\kappa\epsilon
m}\bar\chi\bar\sigma^m\psi^{1}_m \phi_A \plushc \Bigr)
 + 3{\epsilon^2
m^2\over \kappa^2} \ . \label{AdSLG}
\eea

In this expression, $\kappa$ denotes Newton's constant,
$m = \sqrt{\Lambda^2 + \kappa^2 v^4}$  and $D_m$ is the
full covariant derivative.  The scalar-field gauge-invariant
derivatives are as follows,
\bea
\cD_m \phi_A &=& \partial_m \phi_A - m\sqrt{1 - \epsilon}A_m \nonumber \\
\cD_m \phi_B &=& \partial_m \phi_B - m\sqrt{1 + \epsilon}B_m \ ,
\eea
while the supercovariant derivatives take the form
\bea
\hat \cD_m\phi_A&=& \partial_m\phi_A - m\sqrt{1 - \epsilon}  A_m  -
{\kappa\over{2}}(\psi^1_m\chi + \bar\psi^1_m\bar\chi)  \nonumber\\
\hat \cD_m\phi_B&=& \partial_m\phi_B - m\sqrt{1 + \epsilon}  B_m + \I
{\kappa\over{2}}(\psi^1_m\chi - \bar\psi^1_m\bar\chi) \nonumber\\
\hat A_{mn}&=& A_{mn} + {\kappa\over 2}\sqrt{1 + \epsilon}(\psi^2_{
[m}\psi^1_{n] } + \bar\psi^2_{ [m}\bar\psi^1_{n] }) \nonumber \\
& &- \sqrt{1 - \epsilon}{\kappa\over2\sqrt{2}}(\bar\lambda\bar\sigma_{
[n}\psi^1_{m ]} + \bar\psi^1_{ [m}\bar\sigma_{n] }\lambda)  \nonumber
\\
\hat B_{mn}&=& B_{mn} - \I {\kappa\over 2}\sqrt{1 - \epsilon}(\psi^2_{
[m}\psi^1_{n] } - \bar\psi^2_{ [m}\bar\psi^1_{n] }) \nonumber \\
& &+ \I\sqrt{1 + \epsilon}{\kappa\over2\sqrt{2}}(\bar\lambda\bar\sigma_{
[n}\psi^1_{m ]} - \bar\psi^1_{ [m}\bar\sigma_{n] }\lambda)  \ .
\eea

This Lagrangian is invariant (to lowest order in the fields)
under the following supersymmetry transformations,
\bea
\delta e^a_m &\ =\ &\I \,  \kappa \eta^i \sigma^a \overline \psi_{m i} +
   \I \,  \kappa \bar\eta_i \bar\sigma^a \psi_{m}^i \nonumber \\
\delta_{\eta} \psi^1_m &=& {2\over\kappa}D_m\eta^1 + \I \,  {\epsilon
m\over \kappa} \sigma_m \bar\eta^1 \nonumber \\
\delta_\eta A_m &=& \sqrt{1 + \epsilon}\epsilon_{ij} (\psi_m^i\eta^j +
\bar\psi_m^i\bar\eta^j) + \sqrt{1 - \epsilon}
{1\over\sqrt{2}}
(\bar\eta^1\bar\sigma_m\lambda + \bar\lambda\bar\sigma_m\eta^1)
\nonumber \\
\delta_\eta B_m &=& \sqrt{1 - \epsilon}\epsilon_{ij} (-\I \,
\psi_m^i\eta^j + \I \, \bar\psi_m^i\bar\eta^j) +
\sqrt{1 + \epsilon}{\I\over\sqrt{2}}
(\bar\eta^1\bar\sigma_m\lambda - \bar\lambda\bar\sigma_m\eta^1)
\nonumber\\
\delta_\eta \lambda &=& \I \, \sqrt{1 - \epsilon}{1\over\sqrt{2}}\hat
A_{mn}\sigma^{mn}\eta^1 + \sqrt{1 + \epsilon}
{1\over\sqrt{2}}\hat B_{mn}\sigma^{mn}\eta^1 \nonumber \\
& & + \sqrt{2} \, \I \, \epsilon m  \, \phi_A\eta^1
 -\I \, \sqrt{2} \,  {m\over \kappa} \sqrt{1 - \epsilon^2} \eta^2
\nonumber \\
\delta_{\eta} \chi &=& \I \, \sigma^m\bar\eta^1 \hat\cD_m\phi_A -
\sigma^m\bar\eta^1 \hat\cD_m\phi_B
 - 2 \, \epsilon m \,  \phi_A\eta^1
+ 2 {m\over \kappa} \sqrt{1 - \epsilon^2} \eta^2 \nonumber \\
\delta_\eta \psi^2_m &=& {2\over\kappa}D_m\eta^2 + \I \,  {m\over
\kappa} \sigma_m \bar\eta^2
-{\I\over 2}\sqrt{1 + \epsilon} \hat H^A_{+mn}\sigma^n\bar\eta^1 -
m\sqrt{1 + \epsilon}A_m\eta^1 \nonumber \\
& & + {1\over 2}\sqrt{1 - \epsilon}\hat H^B_{+mn}\sigma^n\bar\eta^1 +
\sqrt{1 - \epsilon\over 1 + \epsilon}( \partial_m \phi_A -
\I \, \cD_m\phi_B)\eta^1   \nonumber \\
& & -{\kappa\over 2}\sqrt{1 - \epsilon\over 1 + \epsilon}\psi^1_m
(\delta_{\eta^1}\phi_A
 -\I \, \delta_{\eta^1}\phi_B)
-\I \, \epsilon m\sqrt{1 - \epsilon\over 1 + \epsilon}  \phi_A
\sigma_m \bar\eta^1 \nonumber \\
\delta_\eta \phi_A &=& \chi\eta^1 + \bar\chi\bar\eta^1 \nonumber \\
\delta_\eta \phi_B &=& -\I \, \chi\eta^1+ \I \, \bar\chi\bar\eta^1
\label{AdST}
\eea

This result holds to leading order, that is, up to and
including terms in the transformations that are linear
in the fields.  Note that this representation is irreducible
in the sense that there are no subsets of fields that
transform only into themselves under the supersymmetry
transformations. The Lagrangian (\ref{AdSLG}) describes
the spontaneous breaking of $N=2$ supersymmetry in AdS
space.  It has $N=2$ supersymmetry and a local $U(1)$
gauge symmetry.  In unitary gauge, it reduces to the
massive $N=1$ Lagrangian of eq.~(\ref{AdSL}).

Let us now consider the dual case with one massive
tensor.  The degree of freedom counting is as in
Figure 1.  Note that the massive $N=1$
``vector'' multiplet now contains a massive antisymmetric
tensor.

The Lagrangian and supersymmetry transformations for
this system can be worked out following the procedures
described above. They can also be derived by dualizing
first the scalar $\phi_B$ and then the vector $B_m$
using the method described in \cite{cremdual}.  The
Lagrangian is given by
\al{
& &e^{-1}\cL \ =\ \nonumber\\
&  & -\ {1 \over 2
\kappa^2} {\cal R} + \epsilon^{m n r
s}
\overline
\psi_{i m}
  \overline \sigma_n D_r \psi^i_s
 - \I \overline \lambda \overline \sigma^m D_m \lambda
 - \I \overline \chi \overline \sigma^m D_m \chi  \nonumber \\
& & -\ {1 \over 4} A_{m n} A^{m n} - {1 \over 4}
\cF^B_{mn}\cF^{Bmn} - {1\over 2}\cD_m \phi_A \cD^m
\phi_A
 + {1\over 2}v^{Bm} v^B_m   \nonumber \\
& & - \ \Bigl( {1\over \sqrt{2}} m \sqrt{1 - \epsilon^2}
\psi^2_m \sigma^m
   \overline \lambda
+ m \sqrt{1 - \epsilon^2}  \I\psi^2_m \sigma^m
   \overline \chi  \nonumber \\
& &+\ \sqrt{2} m \I\lambda\chi
   + {1\over 2}m\chi\chi
 +\ m\psi^2_m \sigma^{m n} \psi^2_n
 -   \epsilon m\psi^1_m \sigma^{m n} \psi^1_n   \nonumber \\
& &+\ {\kappa \over 4}  \epsilon_{i j} \psi^i_m
\psi_{n}^{j}
    (\sqrt{1 + \epsilon}H_{A-}^{m n} + \sqrt{1 - \epsilon}\cF_{-}^{Bm
n})  \nonumber \\
& & +\ {\kappa \over {2}}  \chi \sigma^m
    \overline \sigma^n \psi^1_m (\cD_n \phi_A + \I v_n^B)
 \nonumber \\
& &+\ {\kappa \over 2 \sqrt{2}}  \overline \lambda
    \overline \sigma_m \psi^1_n (\sqrt{1 - \epsilon}H_{A+}^{m n} -
\sqrt{1 + \epsilon}\cF_{+}^{Bm n})
    \nonumber \\
& & +\  {\kappa \over {2}}
    \epsilon^{m n r s}\sqrt{1 - \epsilon\over 1 +
\epsilon}
    \overline \psi_{m 2} \overline \sigma_n \psi^1_r
    (\partial_s \phi_A + \I v_s^B)
    \nonumber \\
& & -\ {\kappa \over {2}}m
    \epsilon^{m n r s}
    \overline \psi_{m 2} \overline \sigma_n \psi^1_r
    \sqrt{1 + \epsilon}A_s
     \nonumber \\
& & -\ 2\kappa\epsilon m \sqrt{1 - \epsilon\over
1 + \epsilon} \bar\psi_{m 2}
\bar\sigma^{mn}\bar\psi_{n 1}\phi_A
+ {\kappa\epsilon m \over
\sqrt{2}}\bar\lambda\bar\sigma^m\psi^{1}_m \phi_A
  \nonumber \\
& & +\   \I {\kappa\epsilon
m}\bar\chi\bar\sigma^m\psi^{1}_m \phi_A \plushc \Bigr)
 + 3{\epsilon^2
m^2\over \kappa^2} \ . \label{lagr}
}
where
\al{
\cD_m \phi_A &=& \partial_m \phi_A - m\sqrt{1 - \epsilon}A_m \nonumber \\
\cF^B_{mn} &=& \partial_{[m }B_{n]} - m\sqrt{1 + \epsilon} B_{mn}
}
and
\al{
\hat \cD_m\phi_A&=& \partial_m\phi_A - m\sqrt{1 - \epsilon}  A_m  -
{\kappa\over{2}}(\psi^1_m\chi + \bar\psi^1_m\bar\chi)  \nonumber\\
\hat v_m &=& v_m - \Bigl(\, \I\kappa \psi^{1}_n\sigma_{m}{}^n\chi
                 -{\I\kappa\over
2}\sqrt{1 - \epsilon\over 1 +
\epsilon}\epsilon_{m}{}^{nrs}\psi_n^{1}\sigma_r\bar\psi_s^{2}
\plushc \Bigr)\nonumber \\
\hat A_{mn}&=& A_{mn} + {\kappa\over 2}\sqrt{1 + \epsilon}(\psi^2_{
[m}\psi^1_{n] } + \bar\psi^2_{ [m}\bar\psi^1_{n] }) \nonumber \\
& &- \sqrt{1 - \epsilon}{\kappa\over2\sqrt{2}}(\bar\lambda\bar\sigma_{
[n}\psi^1_{m ]} + \bar\psi^1_{ [m}\bar\sigma_{n] }\lambda)  \nonumber \\
\hat \cF^B_{mn}&=& \cF^B_{mn} + {\kappa\over 2}\sqrt{1 - \epsilon}(\psi^2_{
[m}\psi^1_{n] } + \bar\psi^2_{ [m}\bar\psi^1_{n] }) \nonumber \\
& &+ \sqrt{1 + \epsilon}{\kappa\over2\sqrt{2}}(\bar\lambda\bar\sigma_{
[n}\psi^1_{m ]} + \bar\psi^1_{ [m}\bar\sigma_{n] }\lambda)
  \ .
}
The supersymmetry transformations are as follows:
\al{
\delta e^a_m &\ =\ &\I \,  \kappa \eta^i \sigma^a \overline \psi_{m i} +
   \I \,  \kappa \bar\eta_i \bar\sigma^a \psi_{m}^i \nonumber \\
\delta_{\eta} \psi^1_m &=& {2\over\kappa}D_m\eta^1 + \I \,  {\epsilon
m\over \kappa} \sigma_m \bar\eta^1 \nonumber \\
\delta_\eta A_m &=& \sqrt{1 + \epsilon}\epsilon_{ij} (\psi_m^i\eta^j +
\bar\psi_m^i\bar\eta^j) + \sqrt{1 - \epsilon}
{1\over\sqrt{2}}
(\bar\eta^1\bar\sigma_m\lambda + \bar\lambda\bar\sigma_m\eta^1)
\nonumber \\
\delta_\eta B_m &=& \sqrt{1 - \epsilon}\epsilon_{ij} (
\psi_m^i\eta^j + \bar\psi_m^i\bar\eta^j) -
\sqrt{1 + \epsilon}{1\over\sqrt{2}}
(\bar\eta^1\bar\sigma_m\lambda + \bar\lambda\bar\sigma_m\eta^1)
\nonumber\\
\delta_\eta B_{mn} &=& - 2\eta^1\sigma_{mn}\chi
-\sqrt{1 - \epsilon\over 1 + \epsilon}( \I\,\eta^1\sigma_{[ m}\bar\psi^2_{n ]}
+ \I\,\eta^2\sigma_{[ m}\bar\psi^1_{n ]})
\plushc \nonumber\\
\delta_\eta \lambda &=& \I \, \sqrt{1 - \epsilon}{1\over\sqrt{2}}\hat
A_{mn}\sigma^{mn}\eta^1 - \sqrt{1 + \epsilon}
{\I\over\sqrt{2}}\hat \cF^B_{mn}\sigma^{mn}\eta^1 \nonumber \\
& & + \sqrt{2} \, \I \, \epsilon m  \, \phi_A\eta^1
 -\I \, \sqrt{2} \,  {m\over \kappa} \sqrt{1 - \epsilon^2} \eta^2
\nonumber \\
\delta_{\eta} \chi &=& \I \, \sigma^m\bar\eta^1 \hat\cD_m\phi_A + \hat
v_m\sigma^m\bar\eta^1
 - 2 \, \epsilon m \,  \phi_A\eta^1
+ 2 {m\over \kappa} \sqrt{1 - \epsilon^2} \eta^2 \nonumber \\
\delta_\eta \psi^2_m &=& {2\over\kappa}D_m\eta^2 + \I \,  {m\over
\kappa} \sigma_m \bar\eta^2
-{\I\over 2}\sqrt{1 + \epsilon} \hat H^A_{+mn}\sigma^n\bar\eta^1 -
m\sqrt{1 + \epsilon}A_m\eta^1 \nonumber \\
& &  +
\sqrt{1 - \epsilon\over 1 + \epsilon} \partial_m \phi_A \eta^1   \nonumber \\
& & -{\kappa\over 2}\sqrt{1 - \epsilon\over 1 + \epsilon}\psi^1_m
\delta_{\eta^1}\phi_A
-\I \, \epsilon m\sqrt{1 - \epsilon\over 1 + \epsilon}  \phi_A
\sigma_m \bar\eta^1 \nonumber \\
& & - {\I\over {2}}\sqrt{1 - \epsilon}\hat\cF^B_{+mn}\sigma^n\bar\eta^1
+ \I \sqrt{1 - \epsilon\over 1 + \epsilon} \hat v_m\eta^1 \nonumber \\
\delta_\eta \phi_A &=& \chi\eta^1 + \bar\chi\bar\eta^1 \label{trafo}
}
These fields form an irreducible representation of the $N=2$
algebra.

Each of the two Lagrangians has a full $N=2$ supersymmetry (up
to the appropriate order).  The first supersymmetry is realized
linearly\footnote{In AdS supergravity, the gravitinos undergo
a shift even for linearly realized supersymmetry \cite{deszu};
see (\ref{AdST},\ref{trafo}) and \Ref{gravitino} in appendix {\bf A}.}.
The second is realized nonlinearly: it is spontaneously broken.
In each case, the transformations imply that
\be
\zeta\ =\ {1\over \sqrt3}\,
(\chi - \I \sqrt 2 \lambda)
\ee
does not shift, while
\be
\nu\ =\ {1\over \sqrt3}\,
(\sqrt 2 \chi + \I \lambda )
\ee
does. Therefore $\nu$ is the Goldstone fermion for $N=2$
supersymmetry, spontaneously broken to $N=1$.

We do not know how to unHiggs the other two representations
of the massive spin-3/2 multiplet.  If we use St\"uckelberg
redefinitions as in (\ref{st2}, \ref{st3}), the supersymmetry
transformations are singular as $\epsilon \rightarrow 1$.
If we try to dualize the above representations, the procedure
is thwarted by the bare $\phi_A$ fields in the Lagrangians
and transformation laws.

\section{The Supersymmetry Algebras}

To find the supersymmetry algebras, let us compute the closure
of the first and second supersymmetry transformations to
zeroth order in the fields.  This will allow us to identify
the Goldstone fields associated with any spontaneously broken
bosonic symmetries.

In the case with two scalars (\ref{AdST}), the algebra is as
follows,
\al{
\left[ \delta_{\eta^2}, \delta_{\eta^1} \right] \phi_A&=&2{m\over\kappa}\sqrt{1
- \epsilon^2}
(\eta^1\eta^2+\bar\eta^1\bar\eta^2)   \nn   \\
\left[ \delta_{\eta^2}, \delta_{\eta^1} \right] A_m&=&\sqrt{1 + \epsilon}
{2\over\kappa}\partial_m
(\eta^1\eta^2 + \bar\eta^1\bar\eta^2) \nn  \\
\left[ \delta_{\eta^2}, \delta_{\eta^1} \right] \phi_B&=&-2\I
{m\over\kappa}\sqrt{1 - \epsilon^2}
(\eta^1\eta^2 - \bar\eta^1\bar\eta^2) \nn \\
\left[ \delta_{\eta^2}, \delta_{\eta^1} \right] B_m&=&-\I\sqrt{1 - \epsilon}
{2\over\kappa}\partial_m
(\eta^1\eta^2 - \bar\eta^1\bar\eta^2) \label{alg}
}
{}From these expressions we see that $\phi_A$ and $\phi_B$ are
Goldstone bosons associated with nonlinearly realized $U(1)$
symmetries that are gauged by the vectors $A_m$ and $B_m$.

In the case with one scalar and one antisymmetric tensor,
(\ref{trafo}), the last two lines in (\ref{alg}) are replaced by
\al{
\left[ \,  \delta_{\eta^2}, \,  \delta_{\eta^1} \right] \,  B_m &=&
{2 \over \kappa} \sqrt{1 - \epsilon}\partial_m(\eta^1\eta^2 +
\bar\eta^1\bar\eta^2) +
2\I {m\over \kappa} \sqrt{1 - \epsilon}
(\eta^1\sigma_m\bar\eta^2 - \eta^2\sigma_m\bar\eta^1) \nonumber \\
\left[ \,  \delta_{\eta^2}, \,  \delta_{\eta^1} \right] \,  B_{mn} &=&
-{2 \, \I\over \kappa}\sqrt{1 - \epsilon\over 1 + \epsilon}D_{[m }
(\eta^2\sigma_{ n]}\bar\eta^1 - \eta^1\sigma_{n] }\bar\eta^2)\ .
}
In this case $\phi_A$ and $B_m$ are the Goldstone bosons of
nonlinearly realized $U(1)$'s gauged by $A_m$ and $B_{mn}$.

To find the symmetry algebra, let us consider these algebras
in the limit $\kappa \rightarrow 0$, with fixed $v^2 \neq 0$,
$\Lambda \neq 0$.  This limit corresponds to a fixed AdS
background, in which central charges can be identified.
For the case with two scalars, we find
\bea
\left[\, \delta_{\eta^2},\, \delta_{\eta^1} \,\right] \phi_A &\ =\ &
2v^2 (\eta^1\eta^2 +
\bar\eta_1\bar\eta_2) \nonumber \\
\left[\, \delta_{\eta^2},\, \delta_{\eta^1}\, \right] A_m &=& 0
\label{A} \\
\left[\, \delta_{\eta^2},\, \delta_{\eta^1} \,\right] \phi_B &\ =\ & -2 \I v^2
(\eta^1\eta^2 -
\bar\eta_1\bar\eta_2) \nonumber \\
\left[\, \delta_{\eta^2},\, \delta_{\eta^1} \,\right] B_m &=&
-\sqrt{2} \I v^2 \, {\partial_m\over \Lambda}\,(\eta^1\eta^2 -
\bar\eta_1\bar\eta_2) \ .
\label{B}
\eea
For the case with one scalar and one antisymmetric tensor, the
last two lines are replaced by
\al{
\left[ \,  \delta_{\eta^2}, \,  \delta_{\eta^1} \right] \,  B_m &=&
2\I v^2
(\eta^1\sigma_m\bar\eta^2 - \eta^2\sigma_m\bar\eta^1) \nonumber \\
\left[ \,  \delta_{\eta^2}, \,  \delta_{\eta^1} \right] \,  B_{mn} &=&
-\sqrt{2}\I {v^2\over \Lambda} \nabla_{[m }
(\eta^2\sigma_{ n]}\bar\eta^1 - \eta^1\sigma_{n] }\bar\eta^2) \label{C}
}

Equation (\ref{A})
implies that the real scalar $\phi_A$ is the Goldstone boson
associated with the $U(1)$ generator of the AdS algebra.  (It is
this generator which contracts to a real central charge in flat
space.)  Equation (\ref{B}) ((\ref{C})) indicates that the
scalar $\phi_B$  (vector $B_m$) is the Goldstone boson
associated with a spontaneously-broken $U(1)$ symmetry, one
which is gauged by the vector field $B_m$ (tensor field $B_{mn})$.

These results imply that when $v \ne 0$ and $\Lambda \ne 0$, the full
current algebra is actually $OSp(2,4) \times_s U(1)$, nonlinearly
realized.  The symbol $\times_s$ is a semi-direct product;
it is appropriate because the supersymmetry generators close
into the local  $U(1)$ symmetry.  This construction evades
the AdS generalization of the
Coleman-Mandula/Haag-{\L}opusza\'nski-Sohnius theorem
because the broken supercharges do not exist.  The $OSp(2,4)
\times_s U(1)$ symmetry only exists at the level of the current
algebra; the $U(1)$ symmetry is always spontaneously broken.

The supergravity theories that we have found depend on three
dimensionful parameters: $\kappa$, $\Lambda$, and $v^2$.  Since
we are interested in partial supersymmetry breaking, we shall
keep $v^2 \neq 0$.  We then consider the Lagrangians
(\ref{AdSLG},\ref{lagr}) as a function of $\kappa$ and $\Lambda$ only.
The dimensionless variable $\epsilon = \Lambda / \sqrt{\Lambda^2
+ \kappa^2 v^4}$ is a particularly useful parameter, because the limit
$\epsilon \rightarrow 0$ corresponds to the case of partially
broken $N=2$ supergravity in Minkowski space, while $\epsilon
\rightarrow 1$ approaches the ``massless'' limit of partially
broken supersymmetry in a fixed AdS background.
The full manifold of $N=2$ supergravities, described by the
parameter $\epsilon$, is plotted in Figure 2.  The center
region corresponds to the new AdS supergravities described
above.

\begin{figure}[t]
\epsffile[63 80 500 230]{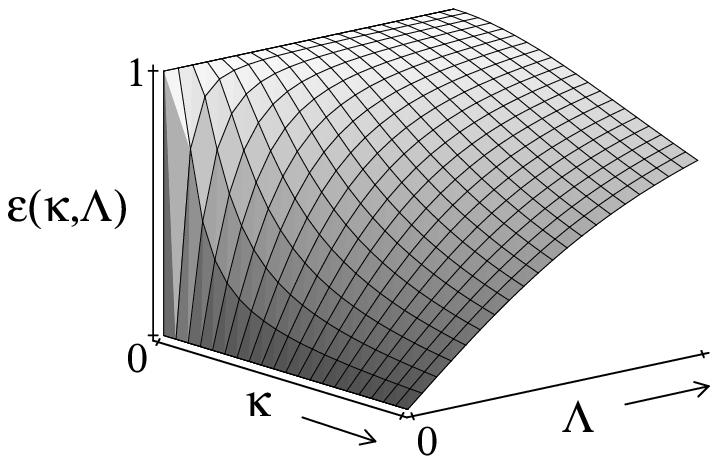}
\caption{The manifold of partially broken $N=2$ supergravity
theories as a function of Newton's constant $\kappa$ and the
cosmological constant $\Lambda$.}
\label{chiral}
\end{figure}

A prominent feature in Figure 2 is the vertical line at
($\kappa =0$, $\Lambda=0$).  This line connects theories
in a Minkowski background ($\epsilon =0$, $\Lambda=0$)
with the ``massless'' limit of theories in a fixed AdS
background ($\kappa =0$, $\epsilon =1$).  The line
suggests that there should be a family of globally
supersymmetric theories in Minkowski space, only one
representative of which ($\epsilon=0$) can be deformed
to a partially broken supergravity theory in a Minkowski
background. In contrast, a continuum of theories ($0 <
\epsilon < 1$) can be deformed to partially broken
supergravity theories in an AdS background.

Indeed, let us consider the limit $\kappa \rightarrow 0$,
$\Lambda \rightarrow 0$ such that $\epsilon$ remains
remains finite.  If we write $\epsilon =\sin(2\theta)$,
we find the following $N=1$ transformations for the case
with two scalars \Ref{AdST}
\al{
  \delta \psi^2_m & = & - {\I \over 2} \cos\theta \overline
    H_{-mn} \sigma^n \overline \eta^1
    - {\I \over 2} \sin\theta
    H_{+mn} \sigma^n \overline \eta^1 +
    \sqrt{2}\partial_m \bar\phi\eta^1 \nn\\
  \delta \cA_m & = & 2 \cos\theta \overline \psi_{m 2} \overline
    \eta^1 + 2 \sin\theta \psi^2_m \eta^1  \nonumber \\
    & & + \sqrt{2} \sin\theta \overline \lambda \overline
    \sigma_m \eta^1 +
    \sqrt{2} \cos\theta \lambda \sigma_m \overline \eta^1 \nn\\
  \delta \chi & = & \I \sqrt{2} \sigma^m \partial_m \phi \overline
    \eta^1 \nn\\
  \delta \lambda & = & {\I \over 2 \sqrt{2}} \sin\theta
    (\overline H_{-mn} \sigma^{mn}) \eta^1
    - {\I \over 2 \sqrt{2}} \cos\theta (H_{+mn} \sigma^{mn}) \eta^1\nn \\
  \delta \phi & = & \sqrt{2} \chi \eta^1 \ .
}
Here, $\phi = (\phi_A + \I \phi_B)/ \sqrt{2}$ and $\cA_m = A_m +
\I B_m$; $\cA_{mn}$ is its corresponding field strength. (The
case with one scalar and one antisymmetric tensor can be obtained
by dualization of $\phi_B$ and $B_m$; it is not presented here.)

The angle $\theta$ can be interpreted in terms of models with a
full $N=2$ multiplet structure \cite{zin, fgp-local}.  In these
models, a necessary ingredient for partial supersymmetry seems
to be presence of at least one vector- and one hyper-multiplet,
as well as the non-existence of a prepotential for the special
K\"ahler manifold \cite{fgp-local} parametrized by the complex
scalars $z_i$ of the $i$ vector multiplets. It was shown in
\cite{special} that such models can always be obtained by a
symplectic transformation from a model with a prepotential.

In \cite{fgp-local} the symplectic vector $\Omega$ for the
special K\"ahler manifold $SU(1,1)/U(1)$ with one complex
scalar $z_1 = z$ takes the form
\eq{
\Omega = \pmatrix{ -{1\over 2} \cr {\I\over 2} \cr \I z \cr z \cr } \ .
}
If we assume that the scalar $z$ acquires a vacuum expectation
value $\langle z \rangle \sim \kappa^{-1}$, and we expand the
supersymmetry transformations \cite{andrian} around this vacuum
expectation value, we find that the angle $\theta$ parametrizes
the symplectic transformation that maps this model with no
prepotential continuously to the case of the so-called ``minimal coupling
models'' \cite{Luciani}:
\eq{
\Omega \rightarrow \pmatrix{\cos\theta & 0 & 0 & -{1\over 2}\sin\theta \cr
                            0 & \cos\theta & -{1\over 2}\sin\theta & 0 \cr
                            0 & 2\sin\theta & \cos\theta & 0 \cr
                            2\sin\theta & 0 & 0 & \cos\theta \cr }
\pmatrix{ -{1\over 2} \cr {\I\over 2} \cr \I z \cr z \cr } \ .
}
Of course, this identification of the angle $\theta$ only holds to
linear order in the fields; at higher order, the model in
\cite{fgp-local} cannot be consistently truncated to our field
content.

\section{Conclusion}

In this paper we have examined the partial breaking of supersymmetry
in anti-de Sitter space.  We have seen that partial breaking in AdS
space can be accomplished using two of four dual representations
of the massive $N=1$ spin-3/2 multiplet.  During the course of this
work, we found new $N=2$ supergravities and new $N=2$ supersymmetry
algebras based on the semi-direct product $OSp(2,4) \times_s U(1)$,
where the $U(1)$ is always nonlinearly realized for finite $\Lambda$.

\section*{Acknowledgements}
We would like to thank A.~Galperin and M.~Porrati for helpful
discussions.  This work was supported by the National Science
Foundation, grant NSF-PHY-9404057.

\newpage

\section{Appendix}

\begin{appendix}
\section{Geometry of AdS space}

Anti-de Sitter space is the space of constant curvature $R<0$.
It has the topology $S^1 \times R^3$ and can be represented as
the hyperboloid
$$
x^Ax^B\eta_{AB} = -{1\over \Lambda^2}\qquad \hbox{ with } \eta_{AB}=\diag(-1\
1\ 1\ 1\ -1)
$$
in flat five-dimensional space. It contains closed time-like curves;
therefore its universal covering space is taken as the physical
space.

In the spirit of nonlinear realizations, the coset spaces of 
anti-de Sitter space and $OSp(1,4)$ are parametrized by the coset elements
\cite{ivsor}
\al{
g(z)&=&\raisebox{.8ex}{$O(3,2)$}/\raisebox{-.8ex}{$O(3,1)$}=\e^{-\I z^mR_m}
\nn\\
G(z,\theta,\bar\theta)&=&{\raisebox{.8ex}{$O(3,2)$}/\raisebox{-.8ex}
{$O(3,1)$}}\cdot
{\raisebox{.8ex}{$OSp(1,4)$}/\raisebox{-.8ex}{$O(3,2)$}} =
 g(z)\e^{\I(1-{1\over 3}\Lambda(\thet\thet + \bar\thet\bar\thet))(\theta
Q^{O(3,2)}
+ \bar\theta\bar Q^{O(3,2)})} \nn
}
with $m \in \{ 0, ..., 3\}$.
Subsequent expressions are facilitated by a
transformation to new bosonic co-ordinates $x^m$
$$
x^m=z^m{\tanh({1\over 2}\Lambda z)\over{1\over 2}\Lambda z}
$$
In these co-ordinates, the vierbein takes the form
$$
{e_m}^a = a(x){\delta_m}^a \ \hbox{ with } a(x)={1\over 1-{\Lambda^2\over
4}x^2}
$$

With the above choice of the fermionic co-ordinate system,
the co-ordinates $\theta$ transform like a Lorentz-spinor
and not like an $O(3,2$) spinor. Therefore, the components
of $OSp(1,4)$ superfields do not transform like $O(3,2)$
fields.

The generators $Q^{O(3,2)}$, however, are $O(3,2)$ spinors. They
can be transformed into $O(3,1)$ spinors $Q$ by shifting
the factor\footnote{Here, $\Lambda(x)$ is the group-theoretical
factor that maps O(3,1)-spinors to O(3,2)-spinors \cite{ivsor};
it is {\it not} the cosmological constant.}
$\Lambda(x)$ to the transformation parameter $\epsilon$:
$$
\e^{\I\epsilon Q^{O(3,2)}}=\e^{\I\eta Q}
$$
Hence the supersymmetry transformation parameter $\eta$ becomes
$x$-dependent \cite{ivsor, suads}:
\eq{
\nabla_m\eta(x)=-\I{\Lambda\over 2}\sigma_m{\bar\eta(x)} \label{xdep}
}

The supersymmetry transformations of the gravitinos (see eqs.
(\ref{AdST}, \ref{trafo})) is also modified by the transition
from $O(3,2)$ to $O(3,1)$ spinors:
\al{
\delta\psi^{O(3,2)}_m&=&\partial_m\epsilon^s(x) \nn\\
\delta(\Lambda(x)\psi_m)&=&\partial_m(\Lambda(x)\eta(x))\nn \\
\delta\psi_m&=&\nabla_m\eta(x) + \Lambda^{-1}(x)(\nabla_m\Lambda(x))\eta(x)
\nn\\
            &=&\nabla_m\eta(x) + \I {\Lambda\over
2}\sigma_m\bar\eta(x)\label{gravitino}
}

The algebra of $OSp(2,4)$ ($i,j \in \{ 1,2 \}$) reads:
\al{
\left[ M_{ab}, M_{bc} \right] &=& -\I\eta_{bb}M_{ac} \nonumber\\
\left[ M_{ab}, R_{c} \right] &=& -\I (\eta_{bc}R_a - \eta_{ac}R_b) \nonumber \\
\left[ R_{a}, R_{b} \right] &=& -\I \Lambda^2 M_{ab}  \nonumber \\
\left[ T^{ij}, Q^k \right]&=&\I \Lambda (\delta^{jk}Q^i - \delta^{ik}Q^j)
\nonumber \\
\{ Q^i_\alpha, \bar Q_{j\dot\beta}\}&=&2\sigma^a_{\alpha\dot\beta}R_a\delta^i_j
\nonumber \\
\{ Q^i_\alpha, Q^{\beta j}\}&=&2\I \Lambda {{\sigma^{ab}}_\alpha}^\beta
M_{ab}\delta^{ij} +
 2\I {\delta_\alpha}^\beta T^{ij}\nonumber \\
\left[ M_{ab}, Q^i \right] &=& -\I\sigma_{ab}Q^i \nonumber \\
\left[ R_{a}, Q^i \right] &=& {1\over 2}\Lambda \sigma_a \bar Q_i \nonumber
}
Here $T^{ij}$ is the hermitian generator of $SO(2)$.
The $N=2$ Minkowski-algebra is recovered in the limit
$\Lambda \rightarrow 0$.  Note that the $N=2$ Poincar\'e
algebra with one central charge is recovered in the
limit $\Lambda T^{ij} = X^{ij}$, with $\Lambda
\rightarrow 0$.

\section{The massive $OSp(1,4)$ spin-1 multiplet}

In section {\bf 2.2}, the unHiggsing of the massive spin-${3\over 2}$
multiplet led to the appearance of a massive spin-${1}$ multiplet
with $\hat E=5/2$. Here, the Lagrangian and transformations for general
$\hat E$ will be presented.

The massive spin-${1}$ multiplet contains the following AdS
representations (see \cite{nicolai}):
$$
D(\hat E,{1\over 2} ) \oplus D(\hat E + {1\over 2}, 1) \oplus
D(\hat E + {1\over 2}, 0) \oplus D(\hat E + 1, {1\over 2}) \quad \hbox{ with }\hat  E \geq
{3\over 2}
$$
The corresponding Lagrangian is:
\al{
\cL & = & - {1 \over 4} v_{mn} v^{mn} - {1\over 2} \partial^m C  \partial_m C
\nonumber \\
& & - \I \bar \lambda \bar \sigma^m \nabla_m \lambda
 - \I \bar \chi \bar \sigma^m \nabla_m \chi \nonumber \\
& & - {1\over 2}\cD_m \phi \cD^m \phi  - {1\over 2}m^2(1 - \epsilon)(1 +
2\epsilon) C^2 \nonumber \\
& & - ({1\over 2}m\lambda\lambda  + {1\over 2}m(1 + \epsilon)\chi\chi \plushc )
\nonumber
}
where $m = (\hat E - {3\over 2})\Lambda \geq 0$ and $\epsilon = {\Lambda /  m}$.
The St\"uckelberg redefiniton
$
\cD_m \phi = \partial_m \phi - m\sqrt{1 + \epsilon}v_m
$
has already been performed.

This Lagrangian is invariant under the supersymmetry transformations:
\al{
\delta_\eta v_m &=&
{1\over\sqrt{1 + {\epsilon\over 2}}}
{1\over\sqrt{2}}(\bar\eta\bar\sigma_m\chi + \bar\chi\bar\sigma_m\eta) \nonumber
\\
& & + \sqrt{1 + \epsilon \over 1 + {\epsilon\over 2}}
{\I\over\sqrt{2}}(\bar\eta\bar\sigma_m\lambda - \bar\lambda\bar\sigma_m\eta)
\nonumber \\
\delta_\eta \lambda &=&   \sqrt{1 + \epsilon \over 1 + {\epsilon\over 2}}
{1\over\sqrt{2}}(v_{mn}\sigma^{mn}\eta - \I {1\over\sqrt{1 + \epsilon}} \cD_m
\phi
\sigma^m\bar\eta) +
{1\over\sqrt{1 + {\epsilon\over 2}}} {1\over\sqrt{2}}(\partial_m C
\sigma^m\bar\eta - \I m (1 - \epsilon)\eta C )\nonumber \\
\delta_\eta \chi &=& {1\over\sqrt{1 + {\epsilon\over 2}}}
{\I\over\sqrt{2}}(v_{mn}\sigma^{mn}\eta
+ \I\sqrt{1 + \epsilon}\cD_m \phi
\sigma^m\bar\eta) -
\sqrt{1 + \epsilon \over 1 + {\epsilon\over 2}}  {\I\over\sqrt{2}}( \partial_m
C \sigma^m\bar\eta + \I m (1 + 2\epsilon)\eta C )\nonumber \\
\delta_\eta C &=& - \sqrt{1 + \epsilon \over 1 + {\epsilon\over 2}}
{1\over\sqrt{2}}
(\eta\chi + \bar\eta\bar\chi) +
{1\over\sqrt{1 + {\epsilon\over 2}}}  {\I\over\sqrt{2}} (\eta\lambda -
\bar\eta\bar\lambda) \nonumber \\
\delta_\eta \phi &=& - \sqrt{1 + \epsilon \over 1 + {\epsilon\over 2}}
{\I\over\sqrt{2}}
(\eta\chi - \bar\eta\bar\chi) -
{1\over\sqrt{1 + {\epsilon\over 2}}}  {1\over\sqrt{2}} (\eta\lambda +
\bar\eta\bar\lambda) \ .
\nonumber
}

In the limit $\hat E\rightarrow {3\over 2}$ ($m \rightarrow 0$)
this Lagrangian reduces to that of a massless spin-1 multiplet
and a chiral multiplet \cite{ivsor}:

\begin{figure}[ht]
\epsffile[60 575 460 720]{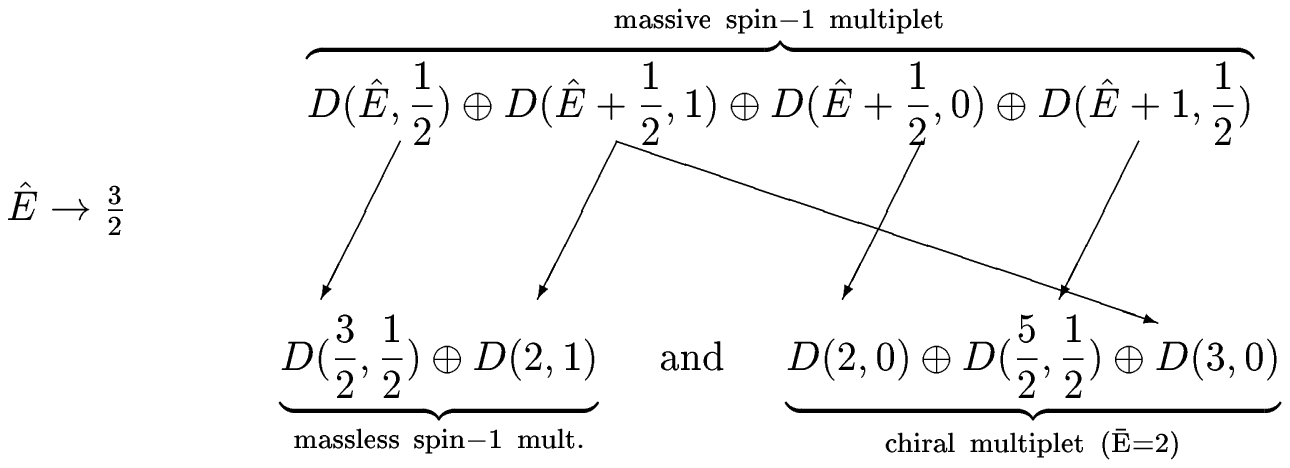}
\end{figure}

It is only this chiral multiplet with $\bar E=2$ that can be dualized
to a linear multiplet in AdS; other chiral multiplets with $\bar E \neq 2$
have bare $\phi$-terms (without derivatives) in its transformations
and cannot be dualized.

The Lagrangian of the dual linear multiplet is

\al{
\cL & = &- {1\over 2}\partial_m C \partial^m C
 +{1\over 2} v_m v^m
- \I \bar \chi \bar \sigma^m \nabla_m \chi \nonumber \\
& & - {1\over 2}\Lambda \chi\chi - {1\over 2}\Lambda \bar\chi\bar\chi
 + \Lambda^2 C^2
}
and its transformations are given by

\al{
\delta_\eta B_{mn} &=& -2\eta\sigma_{mn}\chi -
2\bar\eta\bar\sigma_{mn}\bar\chi\nonumber \\
\delta_\eta C &=& - \chi\eta - \bar\chi\bar\eta \nonumber \\
\delta_\eta\psi &=& -\I\sigma^m\bar\eta \partial_m C
+ 2 \Lambda C \eta  + \sigma^m\bar\eta v_m
\nonumber
}
where $v_m = {1\over 2}\epsilon_{mnrs} \partial^n B^{rs}$.

\end{appendix}

%%%%%%%%%%%%%%%%%%%%%%%%%%%%

%\bibliography{suhi}

\begin{thebibliography}{10}

\bibitem{np}
R.~Altendorfer and J.~Bagger{, Phys. Lett. {\bf B460} (1999) 127.}

\bibitem{cgpads}
S.~Cecotti{, L.~Girardello, and M.~Porrati, Phys. Lett. {\bf B151} (1985) 367.}

\bibitem{nicolai}
H.~Nicolai{, {\it ``Representations of supersymmetry in Anti-de Sitter
  space''}, in {\it ``Supersymmetry and Supergravity '84''}: proceedings of the
  Trieste Spring School on Supersymmetry and Supergravity, World Scientific,
  1984.}

\bibitem{cremdual}
H.~Nicolai and P.~K. Townsend{, Phys. Lett. {\bf B98} (1981) 257.}

\bibitem{deszu}
S.~Deser and B.~Zumino{, Phys. Rev. Lett. {\bf 38} (1977) 1433.}

\bibitem{zin}
S.~Cecotti{, L.~Girardello, and M.~Porrati, Phys. Lett. {\bf B168} (1986) 83;
  \\ Yu.~M.~Zinov'ev, Sov. J. Nucl. Phys. {\bf 46} (1987) 540.}

\bibitem{fgp-local}
S.~Ferrara{, L.~Girardello, and M.~Porrati, Phys. Lett. {\bf B366} (1996) 155.}

\bibitem{special}
B.~Craps{, F.~Roose, W.~Troost, and A.~Van~Proeyen, Nucl. Phys. {\bf B503}
  (1997) 565.}

\bibitem{andrian}
L.~Andrianopoli{, M.~Bertolini, A.~Ceresole, R.~D'Auria, S.~Ferrara, and
  P.~Fr\'e, Nucl. Phys. {\bf B476} (1996) 397; \\ L.~Andrianopoli,
  M.~Bertolini, A.~Ceresole, R.~D'Auria, S.~Ferrara, P.~Fr\'e, and T.~Magri, J.
  Geom. Phys. {\bf 23} (1997) 111.}

\bibitem{Luciani}
J.~F. Luciani{, Nucl. Phys. {\bf B132} (1978) 325; \\ M.~de Roo, J.~W.~van
  Holten, B.~de Wit, and A.~Van Proeyen, Nucl. Phys. {\bf B173} (1980) 175; \\
  B.~de Wit, J.~W.~van Holten, and A.~Van Proeyen, Nucl. Phys. {\bf B184}
  (1981) 77 (E: Nucl. Phys. {\bf B222} (1983) 516).}

\bibitem{ivsor}
E.~A. Ivanov and A.~S. Sorin{, J. Phys. {\bf A} Math. Gen. {\bf 13} (1980)
  1159.}

\bibitem{suads}
P.~Breitenlohner and D.~Z. Freedman{, Annals Phys. {\bf 144} (1982) 249; Phys.
  Lett. {\bf B115} (1982) 197; \\ C.~J.~C.~Burges, D.~Z.~Freedman, S.~Davis,
  and G.~W.~Gibbons, Annals Phys. {\bf 167} (1986) 285.}

\end{thebibliography}
%\bibliographystyle{unsrt}

%%%%%%%%%%%%%%%%%%%%%%%%%%%%

\end{document}